\documentclass[12pt]{article}

\begin{document}


%
%

\title{\textbf {International Spin Physics Symposium 2014 \\ Summary}}

\author{Richard G. Milner \\
Laboratory for Nuclear Science, \\
Massachusetts Institute of Technology, \\
Cambridge, MA 02139, USA}

\maketitle


\begin{abstract}
The Stern-Gerlach experiment and the origin of electron spin are described in 
historical context.  SPIN 2014 occurs on the fortieth anniversary of the first International
High Energy Spin Physics Symposium at Argonne in 1974.  A brief history of the international spin
conference series is presented.
\end{abstract}


\section{Introduction}	
In these brief, concluding remarks to this excellent scientific meeting, I have decided to recount some of the
early origins of spin physics.  I have been motivated to do this by the many young students at this meeting who
may be less familiar with the events almost one hundred years ago on a far-off continent.  Further, I believe that there
are important lessons to be learned for all from an understanding of how the research frontiers are confronted and discoveries are made.  It differs
significantly from the way we learn and teach settled science years later from textbooks in classrooms.

In the early 1920's, the physicist's description of the atom was based on the planetary model of Bohr~\cite{Boh13} and the
quantization of the $z$ component of angular momentum by Sommerfeld~\cite{Som16}.  The electrons orbited the nucleus in
stationary, circular paths at fixed distances from the nucleus.  Electrons could gain or lose energy by jumping from one allowed orbit to another.
The atom had angular momentum $L = n \hbar$, where $n = 1, 2, 3....$ is the principal quantum number.   The electron in the $n = 1$ 
ground state had a magnetic moment of $\frac {e \hbar}{2 m_e c}$, the Bohr magneton.  This model successfully explained the Rydberg formula
deduced from experimental observation of the spectral lines of the hydrogen atom.  While the Bohr-Sommerfeld theory was the accepted description it
was widely recognized that it could not be the final word.  Stern and von Laue are quoted to have declared in 1913: {\it ``If this nonsense of Bohr should
in the end prove to be right, we will quit physics!"}.

\section{The Stern-Gerlach Experiment}

The Stern-Gerlach experiment was carried out in 1922 in Frankfurt, Germany by Otto Stern and Walther Gerlach.  Stern
received a Ph.D. in physical chemistry from the University of Breslau in 1912.  He was the first pupil of Albert Einstein.  Importantly for
our story, Stern was a cigar smoker.  Walther Gerlach received his Ph.D. in experimental physics also in 1912 at the University of T\" ubingen.
Stern had a position at the Johann Wolfgang Goethe University of Frankfurt am Main from 1915 until 1921, when he became a professor at the University of Rostock.  Gerlach held a position at Frankfurt from 1920 until 1925, when he answered a call to become a professor at the University of T\" ubingen.
Max Born was also at the University of Frankfurt from 1919 to 1921, when he left to take a professor position at the University of G\" ottingen.

Otto Stern conceived of the Stern-Gerlach experiment~\cite{Ste21} one cold morning as he lay in his warm bed~\cite{Bre03}.  He was focused on
experimentally demonstrating space quantization.   The idea was to generate an atomic beam of silver atoms from a hot oven.  In the Bohr-Sommerfeld model,
the atoms are assumed to have a magnetic moment which for the ground state $n=1$ would be $\pm \frac{e \hbar}{2 m_e c}$.  The atoms in the oven
at temperature $T$ would have a distribution of velocities given by the Maxwell-Boltzmann distribution so that some would escape a hole in the wall of the oven.  This atomic
beam would pass through an inhomogeneous magnetic field.    The inhomogeneous field will exert a force on the magnetic dipole which should cause the 
silver atomic beam to be split.  The beams are detected some distance beyond the magnet using a photographic plate.  Stern assumed that  $T =1000^\circ$ K, $\frac {\partial B}{\partial z} = 10^4$ gauss cm$^{-1}$, which would produce a separation on the plate of $1.12 \times 10^{-3}$ cm~\cite{Ste21}.

Stern discussed his idea with Born, who was initially unimpressed:  {\it ``I thought always that this space 
quantization was a kind of symbolic expression for something which you don't understand......I tried to persuade Stern that there was no sense in it but then he told me it was worth a try"}.  The experiment took more than a year to realize.  Securing the necessary funding was a major challenge.  Having been convinced of the importance of the experiment, Born gave public lectures
on Einstein and relativity and charged an entrance fee.  Crucially, a check from Harry Goldman (founder of Goldman-Sachs) in New York saved the experiment.

When the experiment was first carried out and Gerlach removed the plate from the vacuum, no sign of the silver was visible~\cite{Bre03}.  Cigar-smoking Stern received the plate from Gerlach and slowly the silver became visible.  Stern's cheap cigars contained sulfur and the smoke interacting with the silver produced
silver sulfide, which is black and easily visible.  To convince skeptics, this effect of the cigar smoke on the silver was reenacted in 2002 and successfully confirmed~\cite{Bre03}.

The published results~\cite{Ger22} were obtained with a magnetic field gradient up to twenty times that assumed in Stern's proposal.  This greatly increased the
separation to about 0.2 mm, which was visible.  It is doubtful that the effect could be seen with the original value of field gradient.  At the time,
the success of the experiment was heralded as a crucial validation of the Bohr-Sommerfeld theory over the classical theory of the atom.  
It showed clearly that spatial quantization exists, a phenomenon that can be accommodated only within a quantum mechanical theory.


\section{The Spinning Electron}

In 1921, A.H. Compton suggested that the electron has a magnetic moment.   In part, this was motivated to explain the observed, mysterious
doubling of atomic states, beyond what was predicted by the Bohr-Sommerfeld quantization rules.  This doubling was known as {\it Mechanische Zweideutigkeit} in German and as {\it duplexity} in English.  In 1925, the Pauli Exclusion Principle was formulated~\cite{Pau25} as: {\it no two electrons can have identical quantum numbers}.

Also in that year, Leiden graduate students Uhlenbeck and Goudsmit first hypothesized~\cite{Uhl25} intrinsic spin as a property of the electron.    This occurred over the strong objections of some prominent physicists but with the support of their advisor, Paul Ehrenfest.  In their Nature letter they write: {\it``It seems that the introduction of the concept of the spinning electron makes it possible throughout to maintain the principle of the successive building up of atoms utilized by Bohr in his general discussion of the relations between spectra, and the natural system of the elements.  Above all, it may be possible to account for the important results arrived at by Pauli without having to assume an unmechanical duality in the binding of the electrons."}  In the succeeding letter in the same journal, Bohr fully agreed. 

The objections to the idea of spin by physicists of the stature of Pauli and Lorentz were not trivial.  The electron was viewed as having a classical radius $r_e = \frac{1}{4 \pi \epsilon_0} \cdot \frac{e^2}{m_e c^2} = 2.8 \times 10^{-15}$ m.  If the electron was spinning with orbital angular momentum of 1 Bohr magneton, then the velocity at the surface of the electron significantly exceeded the speed of light.  A violation of Einstein's theory of relativity was unacceptable.

In 1926, Thomas~\cite{Thomas1926} correctly applied relativistic calculations to spin-orbit coupling in atomic systems and resolved a missing factor of two in the derived {\it g}-values.  Also in 1926, Fermi~\cite{Fermi1926} and Dirac~\cite{Dirac1926} developed the Fermi-Dirac statistics for electrons.  It was immediately applied to describe stellar collapse to a white dwarf~\cite{Fowler1926}, to electrons in metals~\cite{Sommerfeld1927}, and to field electron emission from metals~\cite{Fowler1928}.

In 1928, Dirac developed~\cite{Dirac28} his elegant equation for spin-$\frac{1}{2}$ particles.  In this formulation, solutions are four-component spinors which are interpreted as positive and negative energy states of spin $\pm \frac{1}{2}$ each.   Dirac predicted the existence of the positron, and the theory became the basis for the most precisely tested theory in physics, Quantum Electrodynamics.  By the end of the 1920s, physicists had developed a fundamental understanding of the essential role of electron spin in explaining the electronic structure of the atom.  There exist excellent, personal, historical accounts by Dirac~\cite{Dirac1974}, Uhlenbeck~\cite{Uhlen1976}, and Goudsmit~\cite{Gouds1976} of this period.

In 1927, Wrede, a student of Stern at Hamburg~\cite{Wrede1927}, and Phipps and Taylor at Illinois~\cite{Phipps1927} independently observed the  deflection of atomic hydrogen in a magnetic field gradient.  In  1929, Mott wondered~\cite{Mott1929} if electron spin can be observed directly via the scattering of electrons from atomic nuclei.  Note that in the Appendix to his paper, Mott showed that the Stern-Gerlach experiment cannot be carried out for electrons.  Only in 1942 did Shull {\it et al.} verify~\cite{Shull1943} Mott's prediction in a double scattering experiment which used 400 keV electrons from a Van de Graaf generator.  In the mid-1920s, Heisenberg and Hund postulated the existence of two kinds of molecular hydrogen: {\it orthohydrogen} where the two proton spins are aligned parallel and {\it parahydrogen} where the two proton spins are antiparallel.  By the end of the decade, they had been studied experimentally.  Later, by deflection of orthohydrogen in a magnetic field gradient, Stern and collaborators measured the {\it g}-factor of the proton to be about 2.5 nuclear magnetons~\cite{Estermann1933}, a marked deviation from the Dirac value for a pointlike spin-$\frac{1}{2}$ particle, and the first hint of its internal structure.

In the 1930s, Rabi and collaborators (inc. N. Ramsey and J. Zacharias) using molecular beams in a weak magnetic field measured the magnetic moments and nuclear spins of hydrogen, deuterium, and heavier nuclei~\cite{Kellogg1939}.

By the end of the 1940s, the nuclear shell model had been established~\cite{Mayer1948}.  This explained the properties and structure of atomic nuclei and underscored the essential role of proton and neutron spin.  A key aspect was the strong role of spin-orbit coupling, which was suggested to Goeppert-Mayer by a question from Fermi.

\section{The International Spin Physics Community}

By the middle of the twentieth century, the intrinsic spin of subatomic particles was a cornerstone of the physicist's theoretical understanding of the fundamental structure of matter.  However, spin as an experimental tool became a 
reality only in the 1950's, when a number of seminal experiments were carried out using spin. In 1956, Lee and Yang pointed out that parity should be violated in the weak interaction~\cite{YangLee1956}. 
Shortly afterwards, in 1956, Wu and collaborators observed~\cite{Wu1957} parity violation in aligned $^{56}$Co. In 1958, it was shown experimentally~\cite{Goldhaber1958} using polarization techniques that the neutrino has negative helicity.  1959, the Thomas-Bargmann-Michel-Telegdi equation describing the spin precession of an electron in an external electromagnetic field was derived~\cite{BMT1959}.

In the 1960s, the discovery of pointlike constituents in the proton at SLAC using deep inelastic scattering (DIS) profoundly affected our understanding of the fundamental structure of matter.  A key determination that these constituents had spin-$\frac{1}{2}$ led to their identification as the quarks of SU(3) symmetry.  Important sum rules related to spin-dependent DIS were derived by Bjorken~\cite{Bjorken1966} and by Ellis and Jaffe~\cite{Ellis1974}.  

During this period, the international spin community grew significantly in size to become the active, subfield of international physics we have to-day.  Beginning in 1960 at Basel, symposia on polarization phenomena in nuclear reactions were held every 5 years until 1994.  Beginning in 1974 at Argonne, symposia on high energy spin physics were held every 2 years until 1994.  Beginning in 1996 in Amsterdam, the international spin community became unified and a symposium on spin physics has been held every two years since then.  The International Spin Physics Committee was formed to oversee the organization of this biennial symposium which has taken place here in Beijing, China in the past week.
The published proceedings of these meetings form the essential record of the research activities over this time.  In~\cite{Schieck}, there is a complete tabulation of these meetings as well as references to their proceedings.  Further, important conventions at Basel in 1960 for spin-$\frac{1}{2}$ particles~\cite{Basel} and at Madison in 1970 for spin-$1$ particles~\cite{Madison} were established to facilitate consistent discussion of spin observables. 

\section{Conclusion}

The Stern-Gerlach experiment was the right experiment to demonstrate space quantization but was explained by the wrong theory at the time.
We now know that the silver atom has an unpaired electron in the $5s$ shell and that all the other electrons are paired.   The $5s$ electron is in a zero orbital angular momentum state.  Thus, the inhomogeneous magnetic field exerted a force only on the magnetic dipole moment of the unpaired electron.

Two graduate students postulated spin over the strong criticism of senior physicists at the time.  Their advisor fully supported them.  At the time,
the Stern-Gerlach experiment was not connected to spin.  There is no mention of it in Uhlenbeck and Goudsmit's paper.

Neither the Stern-Gerlach experiment nor the origination of electron spin was recognized by the Nobel Prize Committee. In 1943, Stern was awarded the Nobel Prize in physics for the discovery of the magnetic moment of the proton. 

Space quantization and spin are the cornerstone of the physicist's description of the universe.  Consequences include: nuclear magnetic resonance, the shell model of the nucleus, optical pumping, the laser, the Lamb shift, the anomalous magnetic moments of the leptons, digital communication, atomic clocks, and the global positioning system.

I want to end by extending my warm congratulations to Profs. Haiyan Gao and Bo-Qiang Ma and their colleagues.  SPIN 2014 in Beijing has been an outstanding success due to their considerable efforts.  We look forward to SPIN 2016 at the University of Illinois Urbana-Champaign, USA.

\section*{Acknowledgments}

I would like to acknowledge the support of my colleagues on the International Spin Physics Committee.
In particular, I thank Erhard Steffens for his leadership, efficiency and untiring dedication to 
the stewardship of international spin physics.  My research is supported by the US Department
of Energy Office of Nuclear Physics under Contract Number DE-FG02-94ER40818.


\end{document}